# Solar Neutrons Observed from September 4 to 10, 2017 by SEDA-FIB

K. Kamiya[1], K. Koga[1], H. Matsumoto[1], S. Masuda[2], Y. Muraki[2], H. Tajima[2], and S. Shibata[3]

[1] *Space Environment Group, Tsukuba Space Center, JAXA, Tskuba, Ibaraki 305-8505, Japan*
[2] *Institute for Space-Earth Environment, Nagoya University, Chikusa, 464-8601, Japan*
[3] *Engineering Science laboratory, Chubu University, Kasugai, 487-0027, Japan*

E-mails: *kamiya.kohki, or koga.kiyokazu, or matsumoto.haruhisa@jaxa.jp*
*masuda, or muraki, or tajima@isee.nagoya-u.ac.jp*
*shibata@isc.chubu.ac.jp*

**Abstract**

The SEDA-FIB is a detector designed to measure solar neutrons. This solar neutron detector was operated onboard the ISS on July 16, 2009 and March 31, 2018. Eighteen large solar flares were later observed by the GOES satellite in solar active region 12673 that appeared on September 4 and lasted until September 10, 2017, with intensity higher than > M2. In nine of those solar flares, the SEDA-FIB detected clear signals of solar neutrons, along with five minor excesses. Among these events, we focus on two associated with the flares of X2.2 (SOL2017-09-06) and X8.2 (SOL2017-09-10) that share a common feature: a process of accelerating electrons into high energies as clearly recorded by the FERMI-GBM detector. These events may provide us with useful information to elucidate the ion acceleration process. The X8.2 event was a limb flare that proved adequate for fixing the parameters needed to explain the process of particle acceleration into high energies. According to our analysis, the electron acceleration process may possibly be explained by the shock acceleration model. However, we found that it would be difficult to explain the simultaneous acceleration of ions with electrons, unless the ions were preheated prior to their rapid acceleration.

Presentator:shibata@isc.chubu.ac.jp

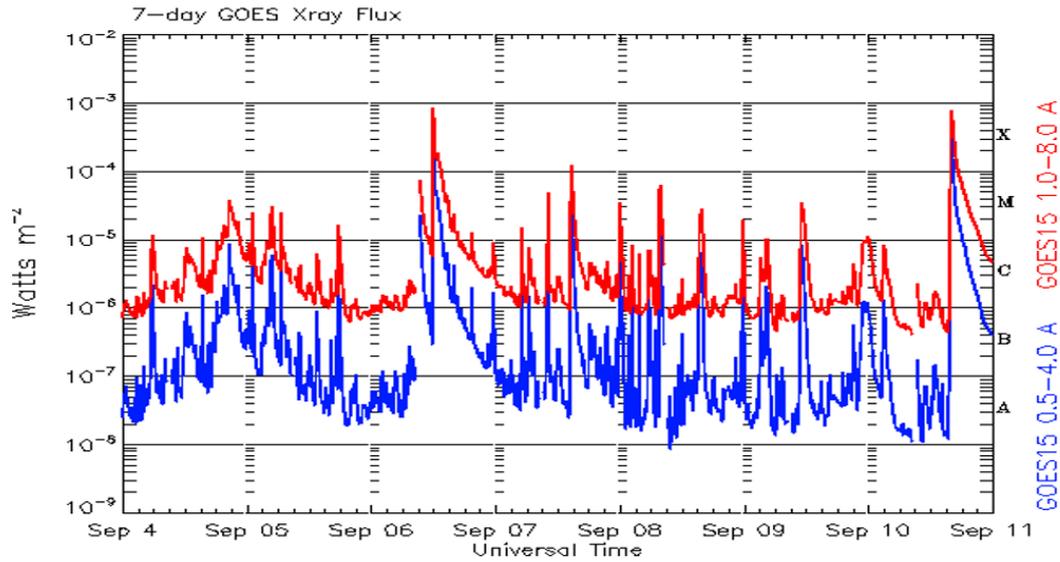

Figure 1. GOES X-ray time profile from September 4 to 11, 2017.

1. Introduction

Many large solar flares were observed in active region (AR) 12673 on the solar surface from September 4 to 10, 2017. **Figure 1** shows the time profile of solar flares that appeared in AR 12673. In association with these flares, high-energy electrons, gamma rays, and neutrons were detected by instruments onboard the satellites of RHESSI [1], Fermi-GBM [2], Fermi-LAT [3], and SEDA-FIB [4-7]. In this paper, we present the energy spectra of solar neutrons detected by the SEDA-FIB. And by comparing the neutron spectrum with the time profile of the process of accelerating electrons into high energies, we also discuss a possible scenario of the acceleration of ions that produced these neutrons.

2. Detection of Solar Neutrons by SEDA-FIB

Figure 1 shows the time profile of soft X-rays recorded by the GOES satellite. We focused on the flares with intensity higher than > M2, and have searched the observed data to determine whether solar neutrons were involved. During this time frame, 18 large flares were recorded by the GOES satellite with intensity > M2. In nine of those solar flares, clear signals of solar neutrons were detected with statistical significance higher than 4.3σ. **Table 1** summarizes those solar neutrons. The neutron events are listed after being renamed from #1 to #9.
2

Let us now explain Table 1 in detail. The sensor of the SEDA-FIB was mounted on the International Space Station (ISS), which orbits Earth every 90 minutes. The ISS also flies over the night region of Earth, so that ~1/3 of the trajectory is masked by Earth's shadow. Thus, during the nighttime of Earth, we could not detect solar neutrons. In the ISS status column of Table 1, "eclipse" represents this situation. Also note that the ISS may pass over the dawn or dusk region, as indicated by the four cases showing the marks of X→O or O→X in the ISS status column. We also present the elevation angle of Earth at the flare time, viewing Earth from the solar horizon. The statuses of the RHESSI and Fermi-GBM satellites are also shown. "X" denotes that the satellite was in the night region of Earth and did not monitor the solar surface.

As you may know, neutrons have mass and thus do not travel at the speed of light in space. This is why time dispersion is associated with solar neutrons. Although we can remove very slow neutrons (i.e., very low-energy neutrons without decay in flight) by using the detector's energy-cut function (at threshold energy of > 35 MeV), we are not completely free from this problem. Therefore, in this paper, we simply set the production time at the peak time of hard X-rays in the energy bands of 50-100 keV or 100-300 keV. Moreover, we sometimes selected the start time at a few minutes earlier than the maximum time. For neutron event #3, the time profile of hard X-rays did not show an impulsive shape, but instead showed a rather extended feature that continued for more than 20 minutes. In such cases, we assumed the production time at a rapid rising time of hard X-rays. In the Comments/start time column, these events were labeled as being "undefined" (i.e., not well determined due to this ambiguity). The neutron start time column lists the assumed start time of these neutrons from the solar surface for each event.

When we derive the statistical significance, an empirical curve was used for the background. When the ISS passes over Northern Canada, we applied an equation of $A*\sin^4(\pi(t-t_0)/45[\text{minutes}]))$ with $A = 3.0$ [events/minute], while defining $A$ as 2.0 [events/minute] for the other paths. The statistical significance is distributed from 9.6σ ~ 6.6σ, except for neutron events #2 and #9 at 4.9σ~6.8σ and 4.3σ~ 4.9σ, respectively. Details on how to select solar neutrons separate from the background (from the ISS body) are provided in other references [4,5,6,7]. The effective area of the sensor is 100 cm$^2$. **Figures 2** and **3** present the differential energy spectra of solar neutrons of the X2.2 flare and the X8.2 flare, respectively. The spectrum shown in Fig. 3 may be well-reproduced by a power law with a power index of $\gamma = -2.0$.

Table 1. List of Solar Neutrons

| No. | Sep. date | GOES max. UT | GOES class | Neutron event # | Observed neutrons | ISS status | RHESSI | FERMI GBM | Comments/ start time | Elevation angle (°) | Neutron start time |
|---|---|---|---|---|---|---|---|---|---|---|---|
| 1 | 4 | 20:33 | M5.5 | 1 | **82** | | O | O | O | 72.2 | 20:28:30 |
| | 4 | 22:14 | M2.1 | | | | X | O | | 75.7 | |
| 2 | 5 | 01:08 | M4.2 | 2 | **62** | | O | O | O | 72.2 | 01:06:50 |
| 3 | 5 | 04:53 | M3.2 | 3 | **110** | | O | O | undefined | 73.8 | ~04:28 |
| 4 | 5 | 06:40 | M3.8 | | | eclipse | X | X | | ------ | |
| 5 | 5 | 17:43 | M2.3 | | | eclipse | X | O | | 65.6 | |
| 6 | 6 | 09:10 | X2.2 | 4 | **102** | | O | O | O | 58.1 | 09:10:00 |
| 7 | 6 | 12:02 | X9.3 | 5 | **130** | X→O | X | X | undefined | 55.9 | ~12:12 |
| | 6 | 15:56 | M2.5 | | | | X | X | | 52.0 | |
| 8 | 7 | 05:02 | M2.4 | | 10+(10) | X→O | O | X | | 45.4 | |
| 9 | 7 | 10:15 | M7.3 | | 20 | | O | X | | ------ | |
| 10 | 7 | 14:36 | X1.3 | 6 | **79** | | X | X | undefined | 40.9 | ~14:35 |
| 11 | 7 | 23:59 | M3.9 | 7 | **64** | | O | X | O | 37.6 | 23:52:00 |
| 12 | 8 | 07:49 | M8.1 | | 33 | | Δ | X | | 33.2 | |
| 13 | 8 | 15:47 | M2.9 | 8 | **117** | | Δ | X | Δ | 22.6 | 15:23:30 |
| 14 | 8 | 23:45 | M2.1 | | 41 | | O | X | | ------ | |
| 15 | 9 | 11:04 | M3.7 | | 27 +(52) | X→O | Δ | X | | 16.3 | |
| 16 | 10 | 16:06 | X8.2 | 9 | **67** | O→X | O | O | O | 3.9 | 15:56:00 |

**Table 1**. List of solar neutrons. From left to right: column 1 [number of flares with intensity > M2]; column 2 [date of flare in September]; column 3 [GOES maximum intensity time]; column 4 [flare intensity defined by GOES]; column 5 [neutron event number]; column 6 [number of neutrons detected (bold numbers correspond to events with high statistics)]; column 7 [ISS status]; columns 8 and 9 [RHESSI and FERMI-GBM locations relative to the Sun]; column 10 [comments on event quality, start time]; column 11 [elevation angle of Earth viewed from the solar horizon]; and column 12 [assumed neutron departure time from the solar surface].

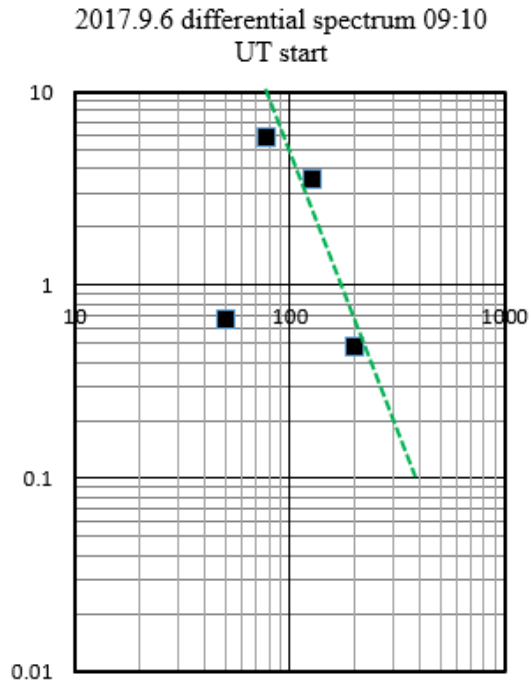
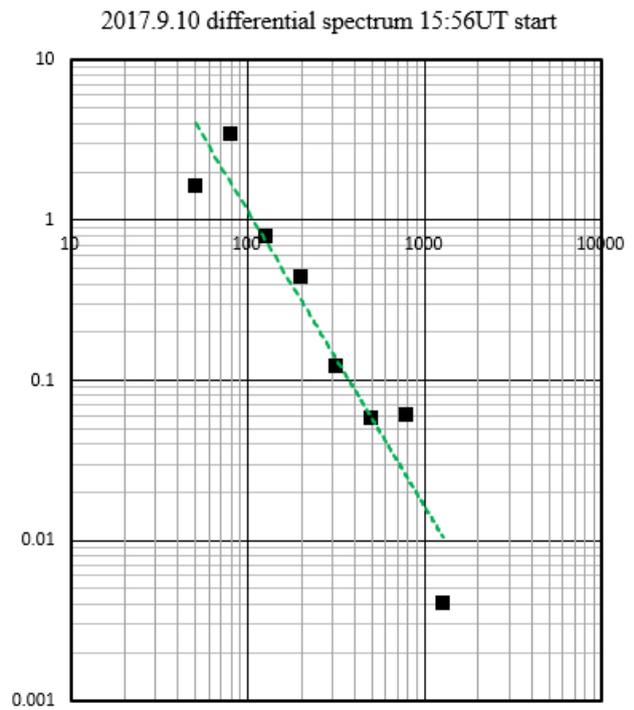

Figure 2. Differential energy spectrum of solar neutrons observed on 2017.9.6. Vertical numbers denote [counts/MeV/detector]. Neutron start time is assumed to be at 09:10 UT.

Figure 3. Differential energy spectrum of solar neutrons observed on 2017.9.10. Vertical numbers denote [counts/MeV/detector]. Neutron start time is assumed to be at 15:56 UT.

## 3. Electron Acceleration Time Profile

One of the remarkable features of active region 12673 may be found in the fact that a very clear acceleration time profile of electrons was obtained. **Figures 4** and **5** show the time profiles of hard X-rays observed by the FERMI-GBM NaI/CsI counter and BGO counter. Figure 4 shows the SOL2017-09-06 event (X2.2); Fig. 5 shows the SOL2017-09-10 event (X8.2). Those hard X-rays were emitted by the electrons accelerated due to the bremsstrahlung process. It is widely accepted when electrons with the power spectrum of $E^{-\delta}$ make bremsstrahlung in a thick target, the observed hard X-ray flux $\gamma$ is related with an equation where $\gamma = \delta - 1$ [8]. The higher part of hard X-rays may be expressed by a power law with $\gamma \approx (4.8 \pm 0.5)$, so electrons may be accelerated with a power index of $\delta \approx (5.8 \pm 0.5)$.

From the observation results, we can see that the energy of hard X-rays increased from 20 keV to 100 keV (for the X2.2 flare) and from 10 keV to 100 keV (for the X8.2 flare) during 10 minutes at the flux of 10 [/s/keV/cm$^2$], respectively. We can also note that the highest point of hard X-rays (soft gamma rays) reached ~300 keV at 09:10 UT (Fig. 4) and at 15:57 UT (Fig. 5).

## 4. A Possible Scenario of High-energy Electron Acceleration

In this section, we will discuss how these electrons were accelerated into high energies. Here we apply a simple shock acceleration model to the events. Observation using the UV telescope of the SDO satellite [9] revealed a candidate magnetic loop (i.e., acceleration site of electrons and possibly ions). The arc length of the loop ($\ell$) may be estimated as being $\ell \simeq 24,000$ km. If electrons with energy travel at nearly the speed of light (~2/3×c) and repeat the back and forth motion within the magnetic loop, the travel time needed to go from one side to the other mirroring point may be estimated as ~0.12 second (**Fig. 6**). Now we assume that Alfvén wave blows continuously from the foot points of the magnetic loops and a shock front may be formed by the collisions of the Alfvén waves. The particles inside the magnetic loop may be also heated up by the Alfvén waves. Then the particles inside the magnetic loop will gain momentum whenever they pass through the shock at every repetition. The momentum gain $\Delta p/p$ may be approximately estimated as ~0.0015 (~300/200,000) per collision. Therefore, initial momentum $P_0$ will increase to the momentum of $P$ after $n$-time collisions. $P$ may be given by $P = 4/3*(1+\Delta p/p)^n \times P_0$.

For example, the number of $n$ for most relativistic particles traveling at nearly the speed of light (~2/3×c) may be given by $300/0.12 \approx 2,500$ times within 300 seconds (five minutes). Therefore, $P \approx 1.002^{2,500} \times P_0 \approx 148 \times P_0$. This implies that the initial electrons with energy of ~10 keV will increase the energy beyond ~a few MeV. According to the calculation taking into account the non-relativistic effect of velocity in the initial stage of acceleration, the momentum will be expected to exceed $\geq 1$ MeV seven minutes later.

According to the results of a recent full-particle Monte Carlo simulation by Hoshino, the low-temperature plasma in the reconnection process will be warmed up to about ~10 keV [10]. And if warm electrons flow from the reconnection site onto the magnetic loop, then the energy spectrum of hard X-rays emitted by the accelerated electrons may thus be possibly explained by the standard shock acceleration model.

## 5. Discussions and Conclusion

When applying the same idea to the ion acceleration process, however, we immediately encounter the difficulty posed by the "wall of particle acceleration". Unless warm ions with momentum higher than a few MeV are prepared before the collisions, it would be quite difficult to accelerate ions into relativistic energies by the shock acceleration mechanism. And in order to explain the observed results (i.e., simultaneous observation of high-energy neutrons and very hard X-rays), such difficulty may therefore exist. This is particularly true when we try to explain the observed results in a unified manner based on a simple shock acceleration model. Consequently, we must either assume the preheating process of ions or introduce a completely different idea such as the DC acceleration mechanism [11]. According to our estimation, such preheating requires more than ~10 minutes for "approach run" time.

**References**

[1] Lin, R.P. et al., Solar Physics, 210 (2002) 3.
[2] Meegan,C. et al., Astrophysical Journal, 702 (2009) 791.
[3] Atwood, W.B. et al., Astrophsyical Journal, 697 (2009) 1071.
[4] Koga, K. et al, Solar Physics, 292 (2017), 115.
[5] Muraki, Y. et al, Advances in Astronmy, 14 (2012) 379304.
[6] Imaida, I et al., *Nucle. Instr. and Meth.*, A421 (1999) 99.
[7] Kamiya, K. et al., EPJ Web of Conference 208 (2019) 14005.
[8] Hudson, H.S., Caanfield, R.C., and Kane, S.R., Solar physics, 60 (1978) 137.
   Kosugi, T, Dennis, B.R., and Kai, K., ApJ, 324 (1988) 1118.
[9] Lemen, J. R. et al., Solar Physics, 275 (2012) 17.
[10] Hoshino, M., Astrophysical Journal Letters, 868 (2018), L18.
[11] Litvinenko, Y. E. and Somov, B. V., Solar Physics 156 (1995) 317, Plasma Atrophysics II, Springer, New York, (2013), ISBN 978-1-4614-4295-0
[12] Fisk, L.A., ApJ, 224 (1978) 1048.
[13] Liu, s., Petrosian, V., and Mason. G., ApJL, 697 (2009),1071.

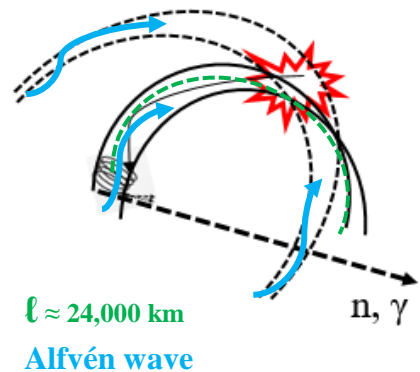

Figure 6. Schematic view of the magnetic loop. The arc length (green dotted line) is estimated as 24,000 km. Ions and electrons inside the magnetic loop repeat the back and forth motion. Alfvén wave rises up from the foot points of the magnetic loops and may collide in the center of the loops, forming the shock front.

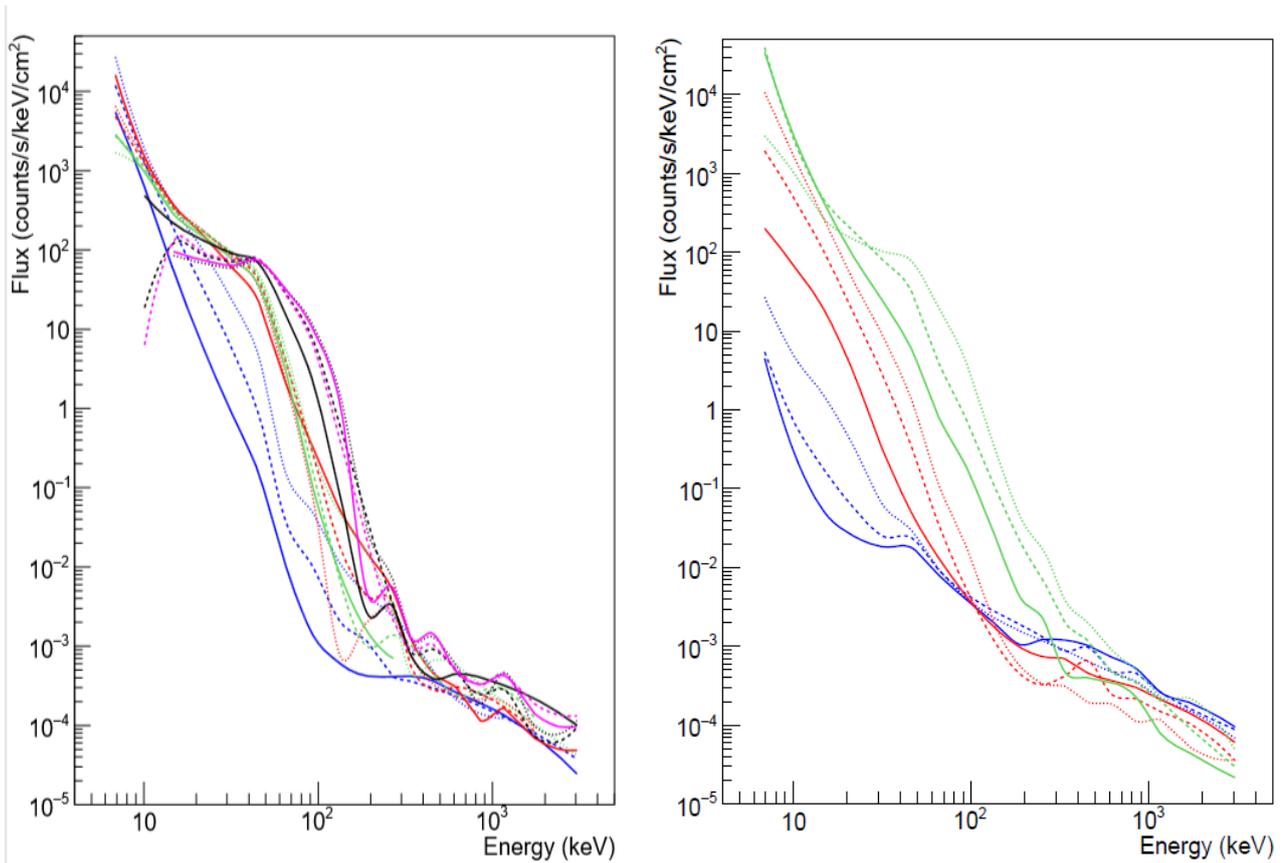

Figure 4. Energy spectrum of hard X-rays observed by FERMI-GBM detectors in association with the September 6, 2017 X2.2 flare. From purple to red, the spectrum was obtained every minute from 09:01:45 UT to 09:10:25 UT. When we draw a horizontal line at the flux of 10 [counts/s/keV/cm2], the corresponding energy of hard X-rays shifts from 20 keV to 100 keV in ~9 minutes. This must be an evidence of electron acceleration with time. The energy spectrum is presented in units of [/sec keV cm$^2$]. The difference of the flux between each energy bin is presented in Figures 4-Append (the next page) and 4B (page 10).

Figure 5. Energy spectrum of hard X-rays observed by FERMI-GBM detectors in association with the September 10, 2017 X8.2 flare. From purple to green, the spectrum was obtained every 100 seconds from 15:44:30 to and 15:58:40 UT. When we draw a horizontal line at the flux of 3 [counts/s/keV/cm$^2$], the corresponding energy of hard X-rays shifts from 7 keV to 100 keV within ~14 minutes. This must be another evidence of electron acceleration with time. The highest energy of hard X-rays reached 400 keV. The difference of the intensity of the flux between each time is presented in Figures 5-Append and 5B.

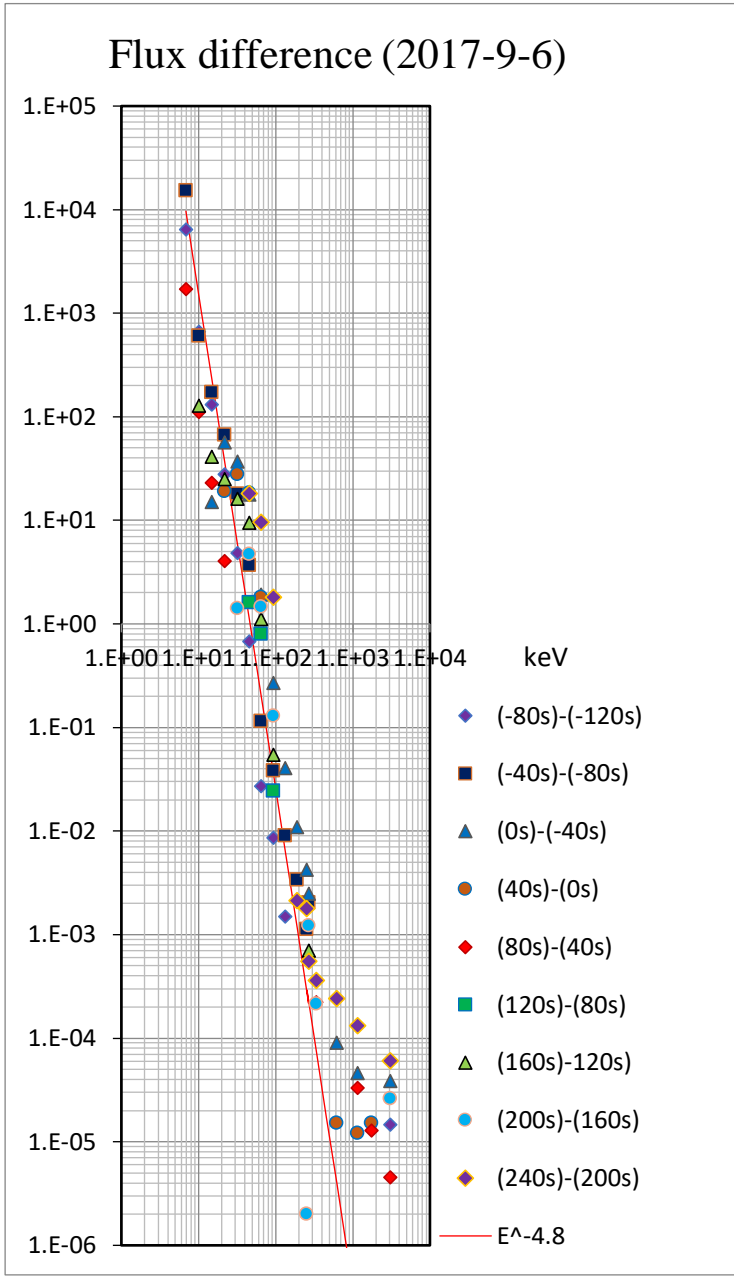
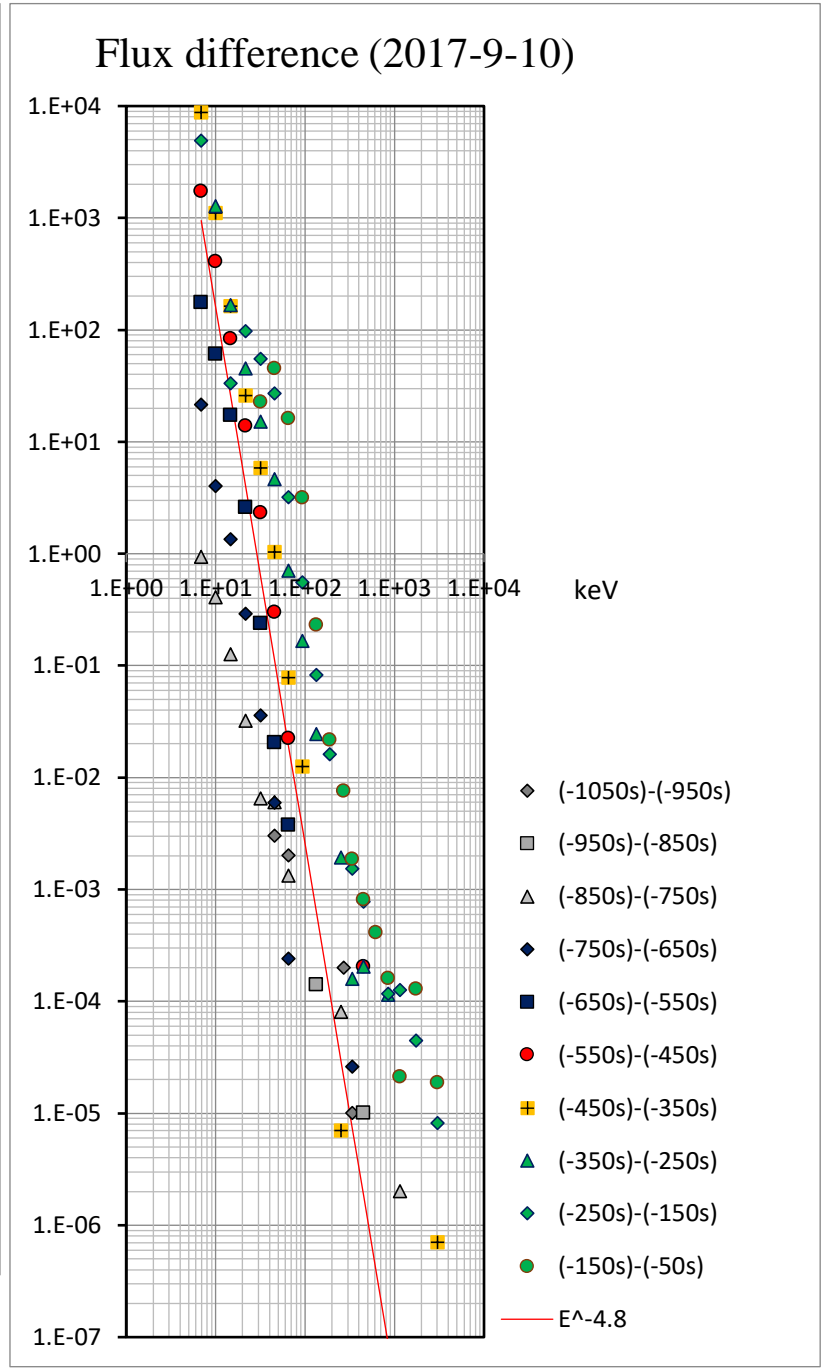

Figure 4-Append.(left) and Figure 5-Append. (right) .

    Above figures present the flux difference between adjacent time.　　In other words, the figures correspond to the increment of hard X-rays within 40 seconds (left side) and 100 seconds (right side) respectively.　　The emission time of hard X-rays was expressed by the time difference from 09:03:45 UT (Figure 4-A) and 15:57:00 UT (Figure 5-A) for September 6th and September 10th flares, respectively.　　In Figure 5, we have plotted the data from -750seconds to 100 sec(=17:58:40UT), while in Figure 5-A the data from -1050 to -750 sec. are included. (Those data are plotted by gray or black.)　　It would be interesting to know that the excess of the flux in the "unit time (Δt)" could be expressed by a universal function approximated by $E^{-4.8\pm0.5}$. This fact suggests us an existence of the shock acceleration region in the flares, i.e., $E_e(t+\Delta t)=E_e(t)(1+R\times\Delta t)$,　　$h\nu(t+\Delta t)=h\nu(1+R\times\Delta t)$,　or　$E_e(t)=E_e(t_o)\mathcal{R}^{\wedge n}$, $n=\Delta t/\Delta t_o$.

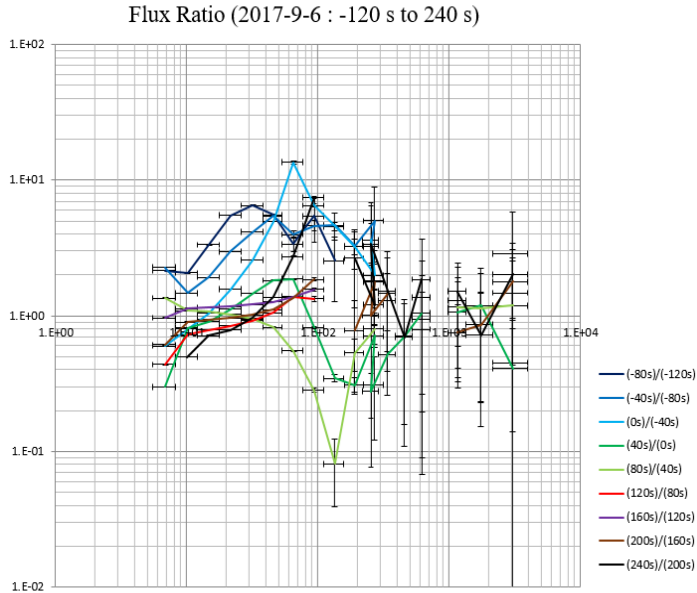

**Figure 4B** (2017-09-07)

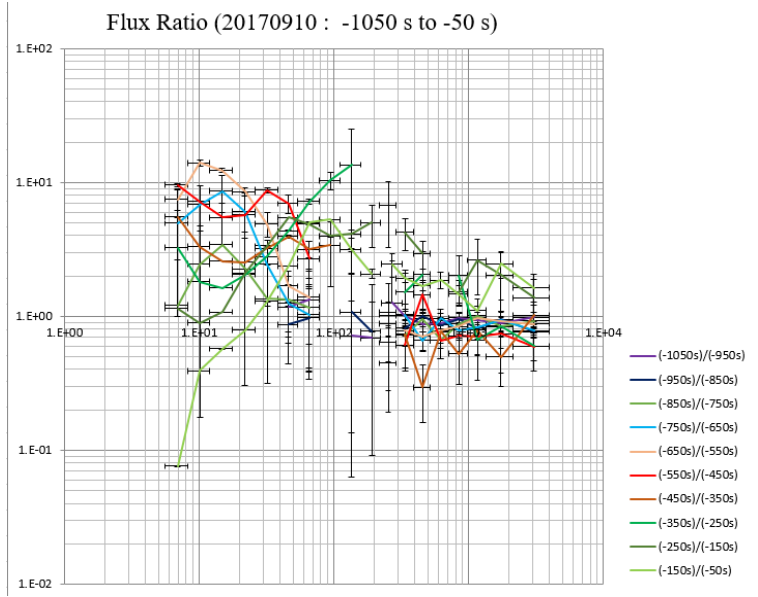

**Figure 5B** (2017-09-10)

Above figures (4B and 5B) present the ratio of the flux between the adjacent time. They are plotted as a function of hard X-ray energies. For example if we compare the two shots between (-550)/(-450sec) (red one) and (-350)/(-250sec) (green one) in Figure 5B, we notice that in the later phase hard X-rays were dominantly produced in such high energy region as 100~200 keV, while in the early stage the increment can be found in rather lower energy region as 10~30 keV. The above fact suggests us that the total number of the warm particles confined in the magnetic bottle, injected from the top of the magnetic loop by the reconnection process, might be not dramatically supplied into the magnetic bottle during the acceleration time.

Then the next question arises; what does work the particle acceleration? Some candidates may be pointed out; the Alfvén wave, the magnet-sonic wave, or the shock waves. Here let us discuss, referring an example of the table tennis. In the table tennis, the ping-pong ball is reflected by the racket. On the other hand this role is acted by the magnetic mirroring at the corner of the *magneti bottle*. The next question is what mechanism did actually work in the acceleration process of particles? One of the candidates may be the Alfvén wave. It arises from the solar surface with the speed of 10~100 km/sec and propagate in the magnetic bottle. The warm particles confined in the magnetic bottle may be resonated with the Alfvén wave via particle-wave interactions and get the momentum. See the ideas of Fisk [12] and Petrosian [13]. This speculation may provide us a clue to resolve the acceleration process of particles in the impulsive flare. Of course, if the shock front is formed in the magnetic loop, moving ~300km/sec inside the magnetic loop, the electron acceleration can be easily explained, as we have already stated in the main text.

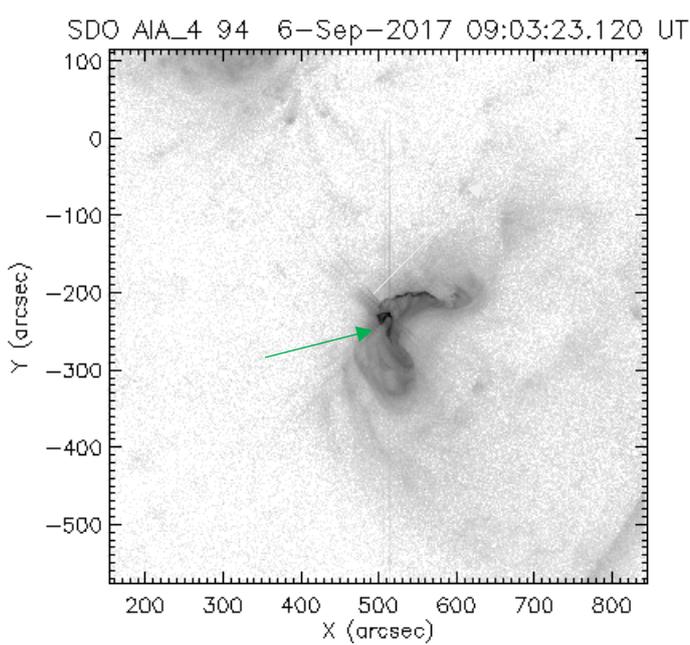

Appendix Figure 1.

The image taken by the 94nm UV telescope on board the SDO satellite. A clear rising magnetic loop is recognized.

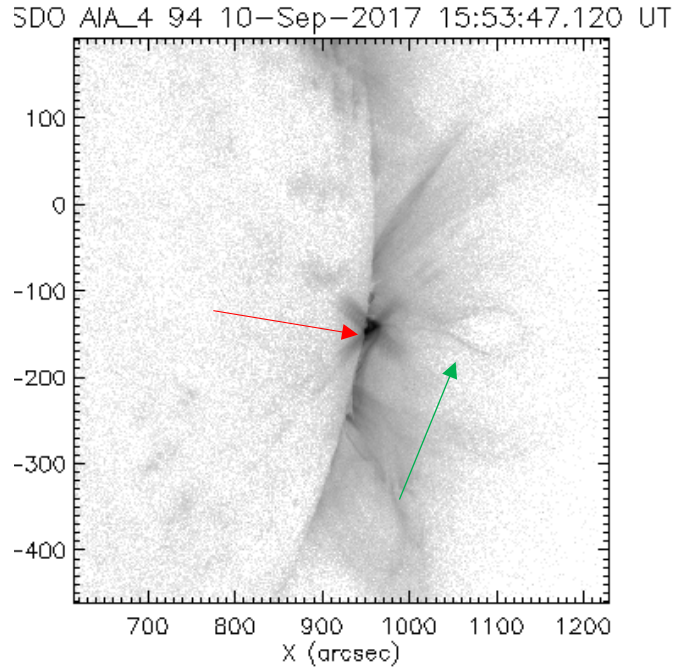

Appendix Figure 2.

The image taken by the 94nm UV telescope on board the SDO satellite.   The emission of the plasmoid is recognized. Hard X-rays were observed near the foot point of the loop (the red arrow).

Appendix Fifures of Solar Neutrons

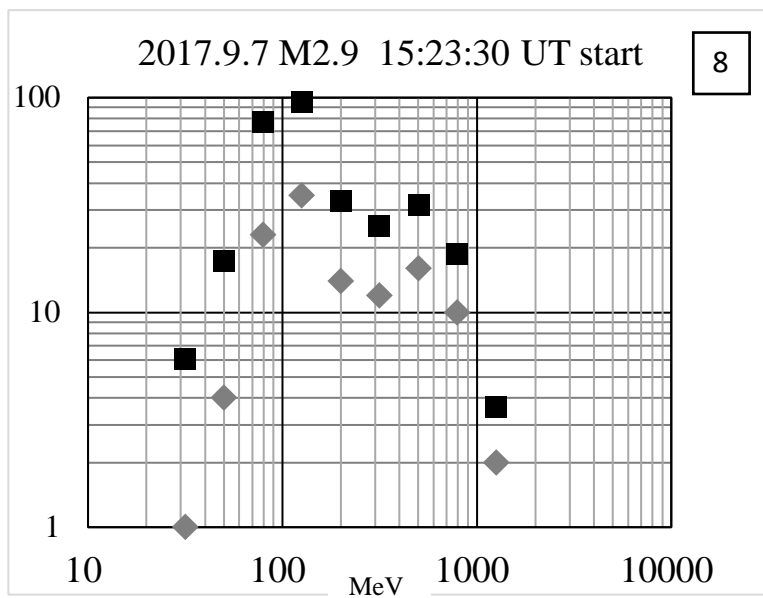

2017.9.7 M2.9  15:23:30 UT start

■ decay corrected
♦ Raw data

#4 → Figure 2
#9 → Figure 3

■ decay corrected
♦ Raw data

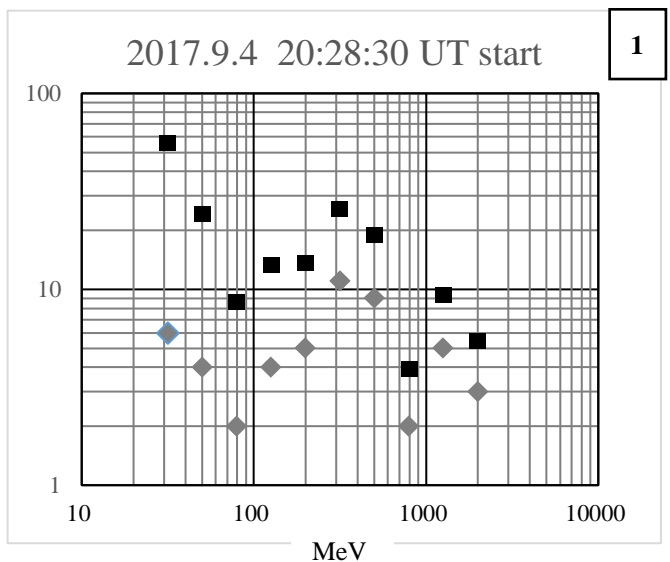

2017.9.4  20:28:30 UT start

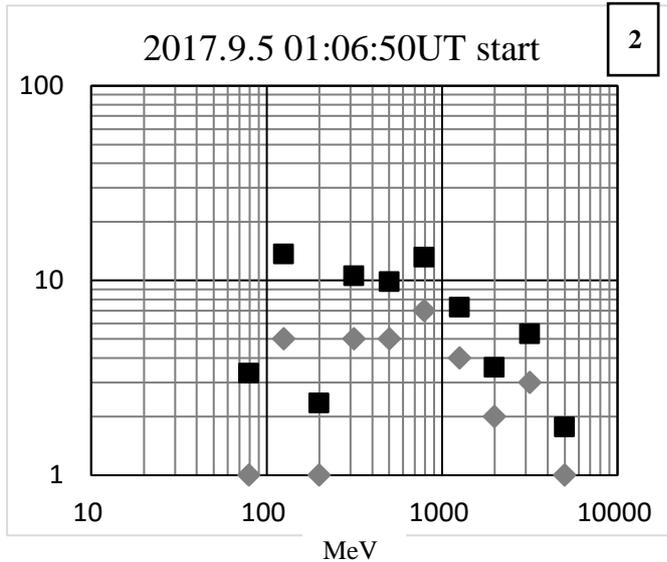

2017.9.5 01:06:50UT start

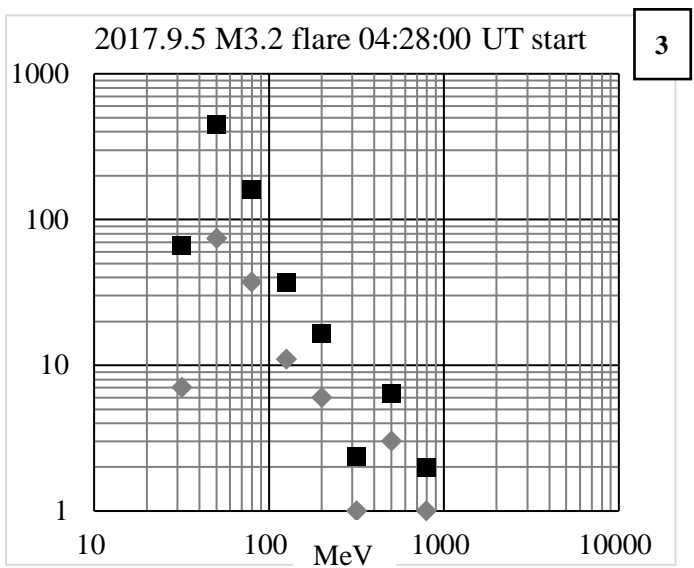

2017.9.5 M3.2 flare 04:28:00 UT start

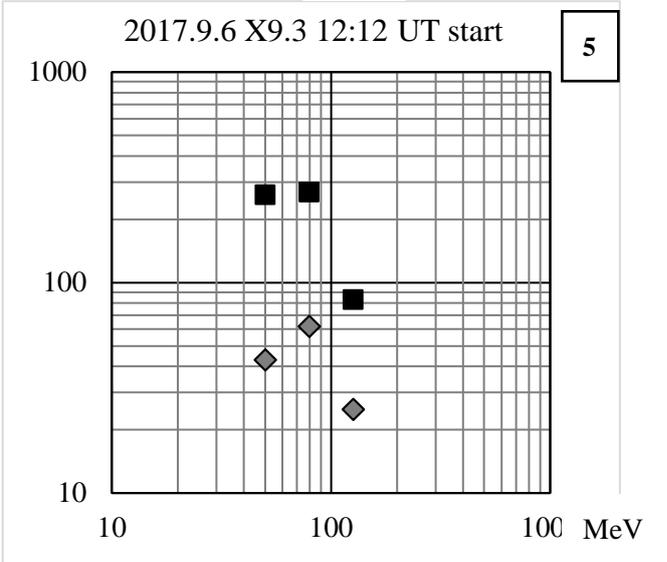

2017.9.6 X9.3 12:12 UT start

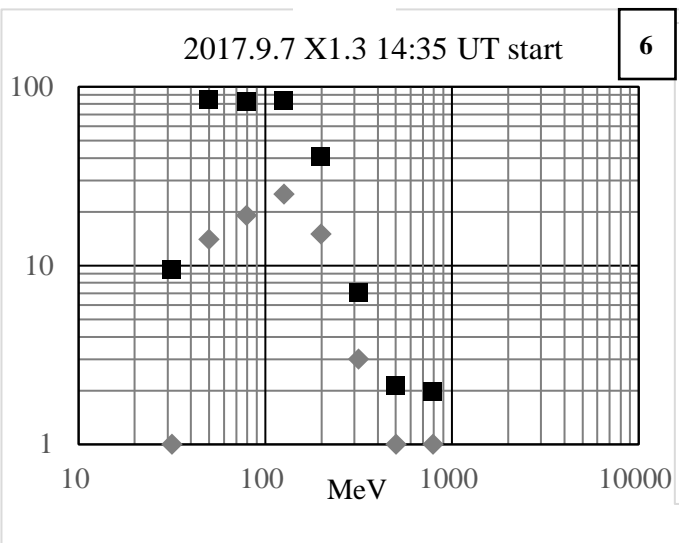

2017.9.7 X1.3 14:35 UT start

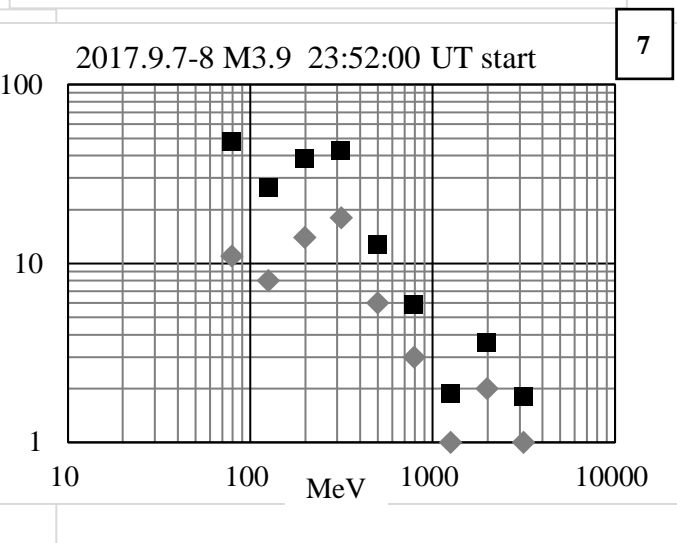

2017.9.7-8 M3.9  23:52:00 UT start